\documentclass[fleqn, secnumarabic, twocolumn, graphics,floatfix,
superscriptaddress,tightenlines, aps, prl]{revtex4}
\usepackage{amssymb}
\usepackage[dvips]{graphicx}
\usepackage{dcolumn}
\usepackage{bm}
\usepackage{color}

\begin{document}

\title{Ultimate photo-induced Kerr rotation achieved in semiconductor
microcavities}
\author{R.V.~Cherbunin}
\affiliation{Spin Optics Laboratory, St-Petersburg State University, 1, Ulianovskaya,
St-Peterbsurg, 198504, Russia}
\author{M.~Vladimirova}
\affiliation{Laboratoire Charles Coulomb, UMR 5221 CNRS/ Universit\'{e} Montpellier 2,
F-34095, Montpellier, France}
\author{K.V.~Kavokin}
\affiliation{Spin Optics Laboratory, St-Petersburg State University, 1, Ulianovskaya,
St-Peterbsurg, 198504, Russia}
\affiliation{Ioffe Physical -Technical Institute of the RAS, 26, Politechnicheskaya,
194021 St-Petersburg, Russia}
\author{A.V.~Mikhailov}
\affiliation{Spin Optics Laboratory, St-Petersburg State University, 1, Ulianovskaya,
St-Peterbsurg, 198504, Russia}
\author{N.E.~Kopteva}
\affiliation{Spin Optics Laboratory, St-Petersburg State University, 1, Ulianovskaya,
St-Peterbsurg, 198504, Russia}
\author{P.G.~Lagoudakis}
\affiliation{Department of Physics \& Astronomy, University of Southampton, Southampton
SO17 1BJ, United Kingdom}
\author{A.V.~Kavokin}
\affiliation{Spin Optics Laboratory, St-Petersburg State University, 1, Ulianovskaya,
St-Peterbsurg, 198504, Russia}
\affiliation{Department of Physics \& Astronomy, University of Southampton, Southampton
SO17 1BJ, United Kingdom}
\date{\today}

\begin{abstract}
Photoinduced Kerr rotation by more than $\pi /2$ radians is demonstrated in
planar quantum well microcavity in the strong coupling regime. This result
is close to the predicted theoretical maximum of $\pi $. It is achieved by
engineering microcavity parameters such that the optical impedance matching
condition is reached at the smallest negative detuning between exciton
resonance and the cavity mode. This ensures the optimum combination of the
exciton induced optical non-linearity and the enhancement of the Kerr angle
by the cavity. Comprehensive analysis of the polarization state of the light
in this regime shows that both renormalization of the exciton energy and the
saturation of the excitonic resonance contribute to the observed optical
nonlinearities.

%
%
\end{abstract}

\maketitle


Semiconductor microcavities elate an increasing interest, due to their
capacity to enhance the light matter interaction. In particular, Kerr
(Faraday) rotation, that is the rotation of the polarization of  light upon
reflection (transmission) from a media characterised by a non-zero
magnetisation projection to the diection of light propagation, can be
increased by orders of magnitude by placing the spin polarised
quasiparticles in a high quality factor planar optical microcavity \cite%
{Hu2008,Cubian2003,Brunetti_2006}. Using this approach, spectacular effects,
such as Kerr rotation by a single electron spin in a quantum dot and Kerr
rotation by nuclear spins were recently observed \cite%
{PhysRevX.4.021004,Giri2013}. Amplification of the electron spin noise, an
effect measured via fluctuations of the Faraday rotation, was achieved by
placing a two-dimensional electron gas in a microcavity \cite%
{PhysRevB.89.205308}. However, experimentally observed polarization rotation
angles in structures with microcavities do not exceed several degrees \cite%
{Brunetti_2006,Giri2012}, though the theoretical limit is $\pi $ radians 
\cite{Kaliteevskii}.

In this work, we demonstrate photo-induced Kerr rotation by
exciton-polaritons in a semiconductor microcavity containing quantum wells
(QWs) in the strong coupling regime by more than $\pi /2$ . This is achieved
by engineering the structure where the exciton polariton mode with strong
spin dependent nonlinear properties is above the impedance matching
condition (IMC). The IMC is characterized by zero reflectivity at the
resonance frequency, which means that the cavity leakage wave and directly
reflected wave at the resonance have equal amplitudes and opposite phases. 
%
%
Above IMC the leakage wave is stronger than the directly reflected wave, so
that photoinduced modifications of the cavity resonance will have the
greatest impact on the reflectivity. To achieve this regime the absorbtion
in the cavity should be sufficiently low. The resonant absorbtion of the
polariton modes makes IMC is difficult to fulfil in microcavities in the
strong coupling regime. Thus, to optimize photoinduced Kerr rotation in a QW
microcavity a trade off between maximum excitonic effects and minimum
absorbtion should be found, by designing the structure where the cavity mode
is below the QW exciton resonance at the IMC. We show, that in such
structure the optical orientation of exciton polaritons by circularly
polarized light induces the splitting between circularly polarized polariton
modes of order of the cavity mode width, sufficient to observe Kerr rotation
angles close to its theoretical maximum of $\pi $. %
%
A careful analysis of the polarization state of the reflected light allows
for the identification and quantitative analysis of the microscopic
mechanisms responsible for the enhanced gyrotropy of microcavities under
optical pumping. Both spin-dependent blue shift of the exciton energy and
the reduction of the exciton oscillator strength are shown to contribute to
the photoinduced Kerr rotation on the equal footing.

\begin{figure}[htb]
\includegraphics[clip,width=1\columnwidth]{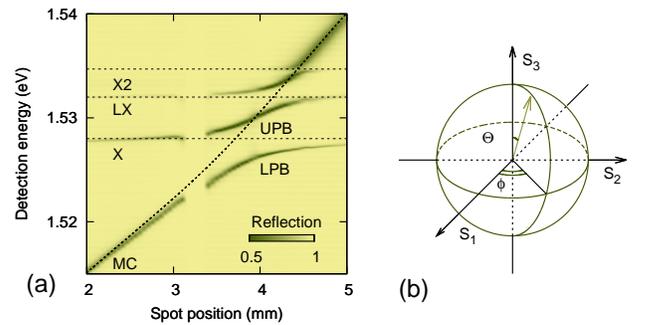}
\caption{(Color online) (a) Reflection spectrum of the structure as a
function of position in the structure plane. (b) Kerr rotation angle $%
\protect\phi$ and ellipticity angle $\Theta$ on Poincare sphere.}
\label{fig1}
\end{figure}

The sample is grown on GaAs substrate and consists of $1\lambda $-cavity
with a $20$-nm quantum well in the center. It is sandwiched between two
Bragg mirrors, the front (rear) mirror consists of $15$ ($25$) pairs of $%
\lambda /4$ AlAs/Al$_{0.1}$Ga$_{0.9}$As layers. This asymmetry is important
to reach IMC. The cavity is grown on a wedge, so that the exciton cavity
detuning can be controlled by choosing the laser spot (of $50$~$\mu $m
radius) position on the sample surface. The structure design is further
detailed in Ref. \cite{Rapaport2001,Lagoudakis2003}. %
%
%
The sample is placed in a closed-cycle optical cryostat at $4.2$~K and
investigated in the backscattering geometry using the time-resolved
pump-probe technique. The gyrotropy in the active layer is created by
optical pumping with circularly polarized pulses of a mode-locked
Ti:Sapphire laser resonant with the upper polariton branch. From the broad
spectrum of the femtosecond laser pulse, pump pulses characterized by the
spectral width of $0.1$~meV are cut by an acousto-optic filter, the
time-integrated pump power varied in the range of $0-20$~mW. Because the
reflection coefficient on the upper polariton branch strongly depends on the
detuning, in the quantitative analysis we use absorbed optical power, rather
than the incident power, assuming that all the power transmitted through the
front mirror is absorbed in the QW. Probe pulses are $20$~meV wide and
linearly polarized in the vertical plane parallel to the crystallographic
axis (110) to minimize the influence of the optical anisotropy of the
sample. The time-integrated power does not exceed $0.1$~mW. The polarization
state of the probe beam is analyzed using an ellipsometer. It consists of
two phase plates (with half-wave and quater-wave retardation) and a linear
polarizer, placed one after another in front of the entrance slit of a $0.5$%
-m spectrometer equipped with a CCD camera. For a given delay between pump
and probe pulses, the spectra of the reflected probe were recorded in six
different polarizations: vertical, horizontal, diagonal, anti-diagonal,
right circular and left circular. These spectra provide the state of
polarization of the probe beam in terms of the Stocks vector components,
which can be mapped to the Poincare sphere (Fig. \ref{fig1}~(b)): $S_{\mu
\nu }=(I_{\mu }-I_{\nu })/(I_{\mu }+I_{\nu })$, where $I_{\mu ,\nu }$ is the
intensity of light components polarized along horizontal ($\mu =H$) and
vertical ($\nu =V$) axes, diagonal (rotated by $\pi /4$) axes: $\mu =D$, $%
\nu =A$, and of the circularly polarized components: $\mu =\sigma +$, $\nu
=\sigma -$. 

Figure 1~(a) shows a color map of the linear reflectivity (in the absence of
the pump) as a function of the photon energy and $x$-coordinate on the
sample surface. The energy of the cavity mode shifts linearly across the
sample. For each position, three dips in the spectrum are observed,
corresponding to the cavity photon mode (MC-mode), heavy hole (X) and light
hole (LX) exciton. Both X and LX modes show anticrossings with the MC mode.
The highest energy anticrossing is due to the second quantized state of the
exciton in the QW (X2). %
Above the anti-crossing point with the upper exciton level, the MC mode
width becomes several times larger due to the inter-band absorption in the
QW and increased losses in the mirrors. These linear reflectivity spectra
are described in the framework of the non-local dielectric response model.
The reflection coefficient of the cavity in the absence of pumping is given
by the function $r(\omega )$, obtained by summation of waves reflected from
all the heterointerfaces \cite{Hu2008}, 
\begin{eqnarray}
r(\omega ) &=&1+\frac{2t_{1}^{2}}{t_{1}^{2}+t_{2}^{2}}\times \frac{i\Gamma }{%
\omega _{c}-\omega -i(\Gamma +\Gamma _{s})-G}, \\
G &=&\sum_{j=1}^{3}\frac{g_{j}^{2}}{\omega _{j}-\omega -i\Gamma _{j}}. 
\nonumber  \label{eq:140306}
\end{eqnarray}%
%
%
Here $\omega _{c}$ is the cavity eigenfrequency, 
$\Gamma $ ($\Gamma _{s}$) are rates of radiative (non-radiative) decay of of
the cavity mode, $g_{j}$ is the strength of coupling between exciton and the
light field (equal to one half of the Rabi frequency \cite%
{kavokin2007microcavities}), $\omega _{j}$ is the exciton resonance
frequency in the quantum well, $\Gamma _{j}$ is the exciton damping rate.
The index $j$ spans over three exciton resonances, X, LX and X2. The mirrors
are characterized by the transmission coefficients $t_{1}$ and $t_{2}$ for
the front and rear mirrors, respectively. The inhomogeneous broadening of
the cavity resonance due to the gradient of the cavity width under the light
spot has been taken into account via convolution of the amplitude reflection
coefficient $r(\omega )$ with the Gaussian distribution of the width $\Gamma
_{inh}$.

The parameters obtained by fitting this model to the linear reflectivity
spectra are summarized in the Table 1. The IMC is achieved in this structure
at the MC mode detuning $\simeq -3$~meV. Note, that due to the inhomogeneous
broadening, the reflectivity does not go to zero at the LPB energy at IMC.
Although LX and X2 polaritons do not induce strong photoinduced Kerr
rotation, it's mandatory to take them into account in the modelling of the
spectra, for the correct description of the polariton states.

\begin{table}[tbp]
\begin{tabular}{l|r|r|r}
\hline
Transition & $\omega_j$ (meV) & $\Gamma_j$ (meV) & $g_j$ (meV) \\ \hline
Heavy exciton, X & 1528.0 & 0.3 & 1.8 \\ \hline
Light exciton, LH & 1532.0 & 0.3 & 1.4 \\ \hline
Upper level exciton, X2 & 1534 & 0.3 & 0.5 \\ \hline
\end{tabular}%
\caption{ Exciton parameters for fitting of the reflection spectrum. Photon
mode parameters: $t1=0.0075$, $t_2=0.00172 $, $\Gamma_s=5 \protect\mu$eV}
\end{table}
%
%

\begin{figure}[t]
\center{\includegraphics[width=0.8\linewidth]{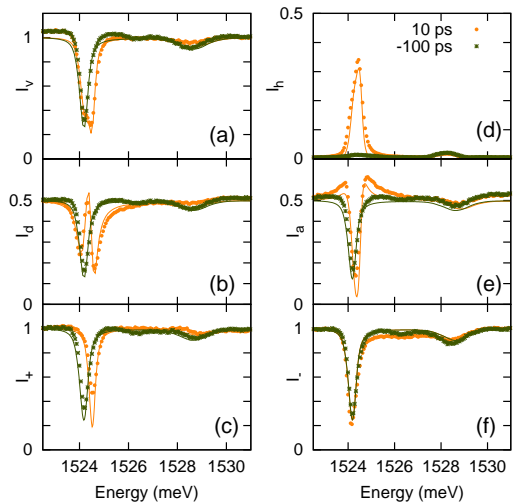}}
\caption{(Color online) Normalized reflectivity spectra detected in presence
of $\protect\sigma^+$ optical pumping for different detection
polarizationsat cavity detuning $\Delta =-3$~meV. Pump power is $P\simeq 1$%
~mW. Spectra taken at $-100$ ps pump-probe delay (they correspond to the
unperturbed system, green symbols) are compared with the photoinduced
spectra measured at $10$ ps pump-probe delay (orange symbols). Lines show
the results of the modelling.}
\label{fig2}
\end{figure}

Fig. \ref{fig2} shows polarization-resolved photoinduced reflectivity
spectra in the vicinity of X resonance, where the strongest photoinduced
effects are observed. The MC mode detuning with respect to the X resonance
is $\Delta =-3$~meV, the pump power is $P\simeq 1$~mW. For each polarization
of detection, two spectra are shown: the spectrum at negative delay, which
is identical to the spectrum of probe in the absence of the pump, and the
spectrum at the pump-probe delay of $10$~ps. In the following we will limit
the discussion to this fixed pump-probe delay. Studies of the polarization
dynamics and relaxation are beyond the scope of this paper and will be
reported elsewhere. Comparing the spectra at negative and positive delays,
one can see a noticeable photoinduced effect in all polarizations, except
counter-circular, with different spectral dependences. In the vertical
linear polarization, $I_{V}$, the lower polariton branch (LPB) splits, and a
signal in the orthogonal horizontal polarization, $I_{H}$, appears. This is
a clear signature of the Kerr rotation. In circular polarization coinciding
with that of the pump (co-circular, $I_{+}$), a spectral shift of the LPB
mode is observed, which is virtually absent in the cross-circular
polarization, $I_{-}$. The upper polariton branch (UPB) is less affected
than the lower polariton branch.

\begin{figure}[tbp]
\center{\includegraphics[width=0.85\linewidth]{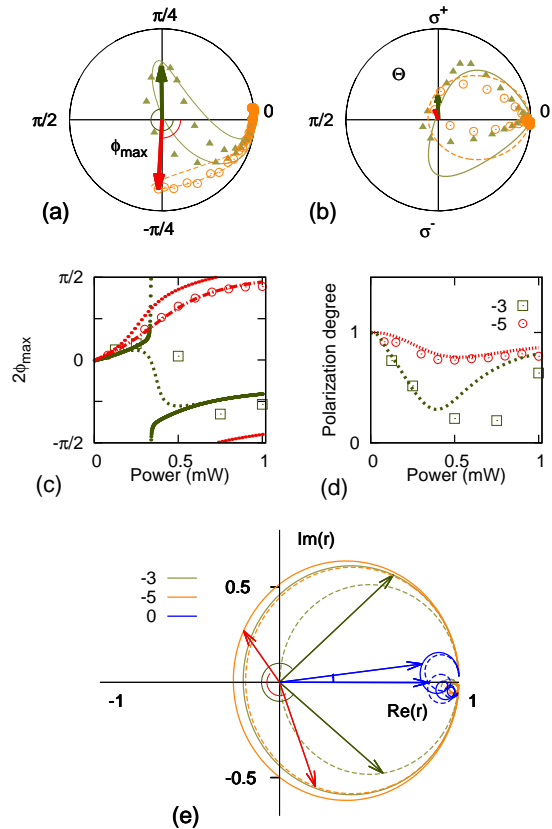}} 
\caption{(Color online) (a) Projection of the polarization of reflected
probe on the equatorial plane of Poincare sphere measured at different probe
energies. Orange circles: $\Delta =-5$~meV, $P=0.4$~mW, green triangles: $%
\Delta =-3$~meV, $P=1$~mW. Lines show results of the fitting. Energy range
is the same as in Fig.~2. (b) The same in the vertical plane of the Poincare
sphere (ellipticity angle). (c) Maximum rotation angle as a function of pump
power for $\Delta =-5$~meV and $\Delta =-3$~meV. Symbols show the
experimental data, dashed lines modelling taking into account inhomogeneous
broadening of the cavity mode, and dotted lines stand for the model without
inhomogeneous broadening. (d) Total polarization degree measured at the
energy of maximum rotation angle for $\Delta =-5$~meV and $\Delta =-3$~meV.
(e) Calculated function $r(\protect\omega )$ on the complex plane for
co-(dashed line) and cross- (solid line) circular polarizations of detection
for $\Delta =-5$~meV (orange), $-3$~meV (green) and $0$~meV (blue). The
angle between two arrows indicates the maximum Kerr rotation angle at each
detuning. }
\label{fig3}
\end{figure}

It is instructive to represent the pump-induced polarization state extracted
from the polarization-resolved measurements in terms of the Stokes vector
hodograph upon variation of the probe energy. In Figs. 3~(a, b) two
different detunings of the photon mode are shown, $\Delta \simeq -3$~meV,
corresponding to IMC, and $\Delta \simeq -5$ meV, that is above IMC. The
energy range is the same as in Fig.~2. Fig.~3~(a) shows the Stocks vector
projection on the equatorial plane. In this representation, the Kerr
rotation angle $\phi $ is readily visualized as the deviation of the Stokes
vector from the abscissae. Arrows indicate the maximum photonduced rotation
angle $\phi _{max}$ measured for each detuning. The projection on the
vertical plane, containing the initial polarization of the incident probe
beam is shown in Fig.~3~(b), its deviation from the vertical axis yields the
ellipticity angle $\theta $. The arrows show the elipticity at the energy of
the maximum Kerr rotation, one can see that it remains small at both values
of $\Delta $.

The maximum values of the Kerr rotation angle measured for each pumping
power are shown by symbols in Fig. 3~(c). The power dependence of Kerr
rotation is monotonic for $\Delta \simeq -5$~meV, $\phi _{max}<\pi /2$. This
corresponds to the fact that the hodograph of the Stokes vector does not
turn around zero (as in Fig. 3(c), red circles) even at the maximum pumping
power. At $\Delta \simeq -3$~meV, it crosses zero at the absorbed power of
about $0.5$~mW (Fig.3(c), green squares). At this power, the Stocks vector
hodograph crosses zero at the equatorial plane, and then reruns back. The
rotation of $\pi /2$ appears as a discontinuity since the angle is given by $%
\phi =atan(S_{DA}/S_{HV})$ and defined between $-\pi /2$ and $\pi /2$. The
further increase of the pumping power allows us reaching $\phi _{max}\sim
3\pi /2$ which the maximum value observed in this work. %

It should be noted that a considerable depolarization of the reflected probe
signal accompanies the Kerr rotation in our experiments. It may be described
by the reduction of the Stokes vector length. The total polarization degree, 
$\rho =\sqrt{S_{HV}^{2}+S_{DA}^{2}+S_{+-}^{2}}$, as a function of absorbed
power is shown in Fig.3~(d). At $\Delta \simeq -5$ meV, the reflected probe
always remains highly polarized, while at $\Delta \simeq -3$ meV a
significant depolarization takes place. At zero and positive detuning, the
system is placed below IMC, and the Kerr rotation angle remains as small as $%
0.1\pi $, even at the maximum applied power (not shown) where the exciton
transition is saturated and the transition from strong to weak coupling
regime occurs.

%

To reproduce the energy, power and detuning dependence of the photoinduced
gyrotropy, we have formulated a simple phenomenological model. We assume
that the effect of the optical pumping can be described in terms of the blue
shift of the X resonance (i) and reduction of coupling with the MC mode in
the co-circular polarization (ii). Both effects are supposed to be linearly
dependent on the absorbed power or, equivalently, on the polariton density.
The importance of the latter contribution in some particular has been
evidenced previously \cite{Huynh2002}, while often it can be neglected \cite%
{Vladimirova2010,Takemura}. Including not only the energy shift but also the
reduction of the coupling is mandatory for the correct description of the
experimental data. The resulting energy of the X state under co-circular
pumping can be written as $\hbar \omega _{X}(P)=\hbar \omega _{X}(0)+\hbar
\delta \omega _{X}P$, and the X-MC coupling parameter as $%
g_{X}(P)=g_{X}(0)+\delta g_{X}P$. Pumping in the counter-circular
polarization is supposed to induce no noticeable variation of the excitonic
parameters at negative detuning, because interaction between polaritons with
opposite spins is negligibly weak in this case \cite%
{Vladimirova2010,Takemura}. The resulting reflection coefficients for right-
($r^{+}$) and left-circular ($r^{-}$) polarizations of light can be
calculated as functions of power using Eq. \ref{eq:140306} with
power-dependent exciton energy and exciton-cavity coupling in the
co-circular polarization. These two reflection coefficients are sufficient
to model the experimental data: the photoinduced reflectivity signal in any
polarization, all the components of the Stocks vector of the reflected
light, and the power-dependence of the Kerr rotation angle. The results of
the fitting of this model results to the measured Stocks vector components
and Kerr rotation angle are shown by solid lines in Fig. 3~(a-d), while the
row spectra in six different polarizations are shown in Fig.~2. Two fitting
parameters $\delta \omega _{X}=0.5$~meV/mW and $\delta g_{X}=0.4$~meV/mW are
used to reproduce the ensemble of the photo induced effects. One can see
that this simple model reproduces quite well the experimental data, namely,
the power dependence of the Kerr rotation and the total polarisation degree.
Note that the inhomogeneous broadening of exciton-polariton modes plays an
important role here. Indeed, dotted lines In Fig. 3~(c) show the Kerr
rotation angle which one could expect for the same cavity parameters but in
the absence of the inhomogeneous broadening. Much sharper increase of the
Kerr angle in the vicinity of $\pi /2$ point could be expected in this case.
Moreover, the depolarisation is entirely due to the inhomogeneous
broadening, and the polarisation degree would not be affected by the pumping
in the homogeneous system.

It is now possible to elucidate the role of the impedance matching condition
in the optically induced gyrotropy. For this purpose we plot in a complex
plane $r^{+}$ calculated at the same power as in Fig.~3~(a) and $r^{-}$
which does not depend on power for $\Delta \simeq 0$, $\Delta \simeq -3$
meV, and $\Delta \simeq -5$ meV, Fig.~3~(e). While the frequency passes
through the cavity resonance, the function $r(\omega )$ makes a circle on
the complex plane, starting from real unity and eventually coming back. The
radius of this circle equals to unity in the system with zero absorbtion,
but in realistic system it is determined by the absorption in the cavity,
and thus is sensitive to the detuning of the MC mode from the X resonance.
The complex reflection coefficient circumvents zero point at $\Delta \simeq
-5$ meV (above IMC), but not at $\Delta \simeq 0$ (below IMC), the
reflectivity turns to zero at $\Delta \simeq -3$ meV, corresponding to the
IMC. Co-circular pumping moves the system towards more negative detunings
due to the photo-induced modification of the excitonic transition energy, so
that the radius of the circle increases (dashed circles), and the
reflectivity vector shifts along this circle. The Kerr rotation angle is
given by $\phi =arg(r_{+}-r_{-})/2$, which is why it's easy to see in this
graphic representation that $\phi $ can't exceed $\pi /2$ below IMC, and can
in principle reach $\pi $ above IMC. The reflectivity vectors corresponding
to the maximum Kerr rotation are indicated by arrows. They are obtained from
the fitting of the data for three different detining values . One can see
that below IMC at zero detuning the rotation is indeed very small. The
maximum rotation is obtained at $\Delta \simeq -3$ meV, where the system is
pushed above IMC by optical pumping. However, at $\Delta \simeq -5$ meV,
further above IMC, the\ Kerr angle is reduced again. This is the consequence
of weaker excitonic effects at stronger negative detunings, and thus smaller
polariton shifts.

The power dependence of upper ($E_{UPB}$) and lower ($E_{LPB}$) polariton
energy shifts measured in co- and counter-circular polarization are shown in
Fig.~4 for two different detunings. One can see that polariton shifts are
essentially linear in density, which justifies the assumptions of the model
(solid lines). At both values of the detuning, LPB experiences the blue
shift, while the shift of $E_{UPB}$ is negative at $\Delta =-5$~meV and
positive at $\Delta =-3$~meV. This behavior is due to the combination of two
excitonic nonlinearities, exciton energy shift and oscillator strength
decrease. In contrast with the experiments of Ref. \cite{Takemura}, where
LPB shifts achieved at the most negative detuning of $\Delta =-2$~meV do not
exceed $0.1$~meV, here stronger LPB shifts indicate higher polariton
densities, and thus the additional mechanism (exciton oscillator strength
decrease) coming into play \cite{Huynh2002}. This regime is beyond the scope
of the present work. Note also, that in the counter-circular configuration
no measurable polariton shift is observed. This is consistent with the
previous measurements, indicating that at strongly negative detunings $%
\Delta <-2$~meV the interaction of polaritons with opposite spins is
negligibly weak.

%
%
%
%
%

\begin{figure}[tbp]
\center{\includegraphics[width=0.85\linewidth]{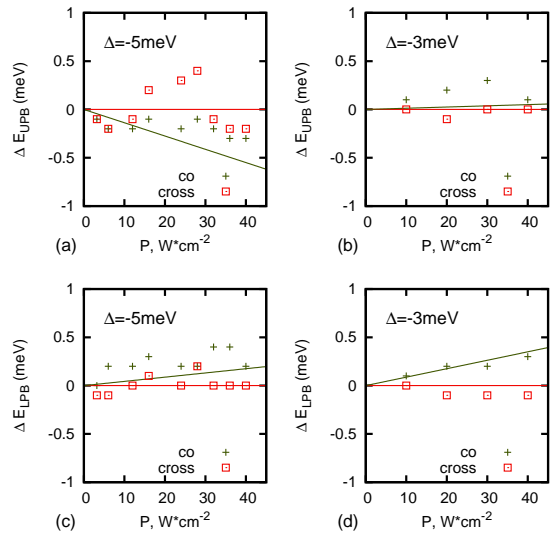}}
\caption{(Color online) Photoinduced energy shifts of the upper and lower
polariton branches in co- (black crosses) and cross (red lines) circular
polarization of detection. (a, c) $\Delta =-5$ ~meV. (b, d) $\Delta =-3$
~meV. Lines are the corresponding energy shifts obtained by the fit.}
\label{fig4}
\end{figure}


In conclusion, we have demonstrated the photoinduced Kerr rotation angle in
semiconductor microcavity of up to $3\pi /2$ radians, close to the
theoretical limit of $\pi $. The conditions which ensure maximum Kerr
rotation constitute a trade off between the needs to maximise photoinduced
excitonic effects (polarization-dependent blue shift and oscillator strength
reduction) close to zero cavity mode detuning, and minimize the absorption
in order to achieve IMC. Understanding of the major role played by IMC is
crucial for the design of specific samples, e.g. suitable for giant
amplification of the optical girotropy. These results open the way for
realisation of fast optical polarisation modulators operating at zero
magnetic field.


\textit{Acknowledgements. }This work was partially supported by the Russian
Ministry of Education and Science (Contract No. 11.G34.31.0067 with SPbSU).
The authors acknowledge Saint-Petersburg State University for a research
grant 11.38.213.2014. RVC thanks RFBR project No. 14-02-31846 for financial
support. MV acknowledge EU INDEX PITN-GA-2011-289968. 
\bibliographystyle{plain}
\bibliography{articles}

\end{document}